Original study

Title: Validation of a new, minimally-invasive, software smartphone device to predict sleep apnea and its severity: transversal study

Authors: Justine Frija[1,2], Juliette Millet[3], Emilie Béquignon[4,5], Ala Covali[6], Guillaume Cathelain[3,] Josselin Houenou[7,8], Hélène Benzaquen[2], Pierre A. Geoffroy[1,9,10], Emmanuel Bacry[11], Mathieu Grajoszex[12], Marie-Pia d'Ortho[1,2]

Affiliations
1- Université Paris Cité, NeuroDiderot, Inserm U1141, F-75019 Paris, France

2- APHP, Hôpital Bichat, Explorations Fonctionnelles et Centre du Sommeil- Département de Physiologie Clinique, Digital Medical Hub académique, F-75018 Paris, France

3- Mitral (Apneal), F-75013 Paris, France

4-Université Paris-Est, INSERM, U955, Equipe 13, CNRS ERL 7000, F-94010, Créteil, France

5-APHP, Service d'Oto-Rhino-Laryngologie et de Chirurgie cervico-faciale, Hôpital Henri Mondor et Centre Hospitalier Intercommunal de Créteil, F-94010, Créteil, France.

6 APHP, Hôpital Henri Mondor, Physiologie Explorations Fonctionnelles, F-94010 Créteil, France

7- NeuroSpin neuroimaging platform, UNIACT lab, Psybrain team, CEA Saclay, Gif sur Yvette, France

8- APHP, Mondor University Hospitals, Psychiatry Dept, INSERM U955 Team 'Translational Neuropsychiatry, Université Paris Est Créteil, Faculté de Santé, Créteil, France

9- Centre ChronoS, GHU Paris - Psychiatry & Neurosciences, 1 rue Cabanis, 75014 Paris, France

10- APHP, Hôpital Bichat Claude Bernard, Département de psychiatrie et d'addictologie, GHU Paris Nord, DMU Neurosciences, F-75018 Paris, France

11- CEREMADE, CNRS-UMR 7534, Université Paris-Dauphine PSL, 75016 Paris, France

12- Digital Medical Hub SAS, Assistance Publique Hôpitaux de Paris AP-HP, Hotel Dieu, Place du Parvis Notre Dame, 75001 Paris, France

Corresponding author :
Justine FRIJA
ORCID ID: 0000-0001-5575-3913
Postal address: Physiologie-Explorations fonctionnelles, Hôpital Bichat-Claude Bernard, 46 rue Henri Huchard, 75018 Paris, France
Fax: +33140258800   Phone: +33140258518        Email: justine.frija@aphp.fr

ORCID IDs for other authors:

Juliette Millet: 0000-0002-1562-6909
Guillaume Cathelain : 0000-0001-7852-8807
Emilie Béquignon : 0000-0002-5193-5776
Ala Covali: 0009-0003-9774-614X
Josselin Houenou: 0000-0003-3166-5606
Hélène Benzaquen : 0009-0008-8185-2606
Emmanuel Bacry : 0000-0001-5997-1942
Pierre A. Geoffroy : 0000-0001-9121-209X
Mathieu Grajoszex
Marie-Pia d'Ortho : 0000-0003-3119-0970

# ABSTRACT:

**Background:** obstructive sleep apnea (OSA) is frequent and responsible for cardiovascular complications and excessive daytime sleepiness. It is underdiagnosed due to the difficulty to access the gold standard for diagnosis, polysomnography (PSG). Alternative methods using smartphone sensors could be useful to increase diagnosis.

**Objective**: assess the performances of Apneal®, an application that records the sound using a smartphone's microphone and movements thanks to a smartphone's accelerometer and gyroscope, to estimate patients' AHI.

**Methods:** monocentric proof-of-concept study with a first manual scoring step, and then automatic detection of respiratory events from the recorded signals using a sequential deep-learning model which was released internally at Apneal® at the end of 2022 (version 0.1 of Apneal® automatic scoring of respiratory events), in adult patients during in-hospital polysomnography.

**Results:** 46 patients (women 34%, mean BMI 28.7 kg/m2) were included.
For AHI>15, sensitivity of manual scoring was 0.91 (95% CI [0.8, 1]), and positive predictive value (PPV) 0.89 (CI 95% [0.76, 0.97]). For AHI > 30, sensitivity was 0.85 (95% CI [0.67, 1]), PPV 0.94 (CI 95% [0.8, 1]). We obtained an AUC-ROC of 0.85 (95% CI [0.69, 0.96]) and an AUC-PR of 0.94 (95% CI [0.84, 0.99]) for the identification of AHI > 15, and AUC-ROC of 0.95 (95% CI [0.86, 0.99]) and AUC-PR of 0.93 (95% CI [0.81, 0.99]) for AHI > 30. The ICC between the AHI estimated manually, and from the PSG is 0.89 (p-value = $6.7 \times 10^{-17}$), Pearson correlation 0.90 (p-value=$1.25 \times 10^{-17}$).
For AHI>15, the automatic Apneal® scorings compared to the PSG scorings led to a sensitivity of 1 (95% CI [0.95, 1]), and a PPV of 0.9 (95% CI [0.8, 0.9]) for the AHI threshold of 15. For the AHI threshold of 30, we obtained a sensitivity of 0.95 (95% CI [0.84, 1]), and a positive predictive value of 0.69 (CI 95% [0.52, 0.85]). Using the smallest threshold to obtain a PPV> 0.9 for AHI > 15, we find a sensitivity of 0.97 (95% CI [0.91, 1]) and PPV of 0.9 (95% CI [0.79, 0.98]); for AHI > 30 sensitivity of 0.57 (95% CI [0.35, 0.78]) and PPV of 0.93 (95% CI [0.75, 1]). The ICC between the AHI estimated, and from the PSG scorings is 0.84 (p-value = $5.4 \times 10^{-11}$) and the Pearson correlation found is 0.87 (p-value = $1.7 \times 10^{-12}$).
Conclusion: manual scoring of smartphone-based signals is possible and accurate compared to PSG-based scorings. Automatic scoring method based on a deep learning model provides promising results. A larger multicentric validation study, involving subjects with different SAHS severity is required to confirm these results.

Trial registration : NCT03803098

# INTRODUCTION

Sleep apnea-hypopnea syndrome (SAHS) affects 15-25% of the population in developed countries[1][2]. It is characterized by repeated interruptions of ventilation during sleep, resulting in sleep disturbance, chronic intermittent hypoxia, and hypercapnia. SAHS triggers cardiovascular diseases and excessive daytime sleepiness, which can result in occupational and road traffic accidents. Proper treatment of SAHS, mostly by continuous positive airway pressure (CPAP), reduces the risk of accidents and injuries[3].

Diagnosis of SAHS is based on a polysomnographic recording (PSG) of respiratory- (chest belt, abdominal belt, nasal flow, nasal pressure, snoring, oxygen saturation), cardiac activity (heart rate, photoplethysmogram, electrocardiogram), brain- (EEG), and motor activities (EMG, actimeter, position sensor). While PSG is considered as the reference to precisely evaluate the SAHS severity of a patient, in clinical practice in adults, a simplified ambulatory polygraph recording (PG) is generally sufficient for the diagnosis of SAHS and records at least the following sensors: abdominal and thoracic belts, nasal scope, microphone, pulse oximeter, position, and oxygen saturation [4]. The metric used to evaluate the severity of SAHS is the Apnea Hypopnea Index (AHI), which is the number of apneas and hypopneas per hour (of sleep). Patients will be considered normal (without SAHS) if their AHI is under 5, as having mild SAHS if their AHI is above 5 and below 15, as having moderate SAHS if their AHI is between 15 and 30, and severe SAHS if their AHI is above 30 [5].

However, SAHS screening and diagnosis are too infrequent, and a large number of apneic patients are unaware of their condition. For example, 80% of apnea sufferers in the USA remain undiagnosed[6]. The main obstacles to diagnosis are patients' awareness, general practitioners' lack of training in this pathology, the cost of diagnosis and treatment, and access to diagnostic tests [7].

For screening purposes in the general population, some questionnaires can be used like the STOP-BANG [8] or the Berlin questionnaire [9]. While these questionnaires are easy to use, their screening results are either too sensitive or do not show enough sensitivity[8,9].

Recently, to obtain more accurate results than questionnaires, alternative signals have been proposed, with devices that offer a simpler installation than a PSG, and are used at patients' homes. For example, some devices use tracheal sounds [10], mandibular movement[11], or movement detected through sleep mattresses[12]. Some of these are validated of OSA diagnosis or screening, and offer multinight-testing, limiting the risk of misdiagnosis [13]. However, all these devices require dedicated hardware.

To overcome this need of a specific device, inducing economic and environmental questions, and to take advantage of the powerful sensors that exist in smartphones, researchers have considered methods to screen SAHS either using smartphones' accelerometers[14] or the sound recorded by a smartphone's microphone[15], and obtained promising but insufficient results. The use of smartphone sensors seems also interesting in reconstructing other vital signs that can be useful to doctors when screening for SAHS, like heart rate[16].

In this paper, we propose a new method that uses no hardware but smartphone's available sensors to estimate patients' AHI, called Apneal®. This application records the sound using a smartphone's microphone and movements thanks to a smartphone's accelerometer and gyroscope. This proof-of-concept study was conducted with a first manual scoring step, and then automatic detection of respiratory events from the recorded signals using a sequential deep-learning model which was released internally at Apneal® at the end of 2022 (version 0.1 of Apneal® automatic scoring of respiratory events).

# METHODS

Study design

We conducted a transversal study at a university hospital sleep center (Centre du sommeil, hôpital Bichat Claude Bernard, Paris, France). Consecutive adult patients referred for videopolysomnography (VPSG) as part of routine care were offered to participate. Other inclusion criteria were ability to understand French, health insurance coverage. Exclusion criteria were known cardiac rhythm disorder, pacemaker, diabetes mellitus, CPAP treatment during the night of the study, or refusal to participate. In this proof-of-concept study, no power calculation was made.

Data acquisition
After obtaining informed consent, patients were equipped with both devices: the polysomnography (PSG) sensors were installed and a smartphone in airplane mode (iPhone 12 mini) was attached to their chest thanks to an adhesive band. The Apneal® application was then switched on. The smartphone was placed with a microphone directed towards the mouth of the patient, and the screen facing up (see Figure 1).

Full VPSG (Alice 6, Philips Respironics) was performed and scored following American Association of Sleep Medicine (AASM) 2012 guidelines by trained technologists and doctors[17]. We named this first scoring of the PSG **PSG-first.** After this first scoring, the scoring of the respiratory events was verified and modified if needed by other trained technologists. We called this second version of the scoring of the PSG **PSG-second.** Both scoring versions can be considered as a reference.

The installation of both device on the patients are visible on Figure 1.

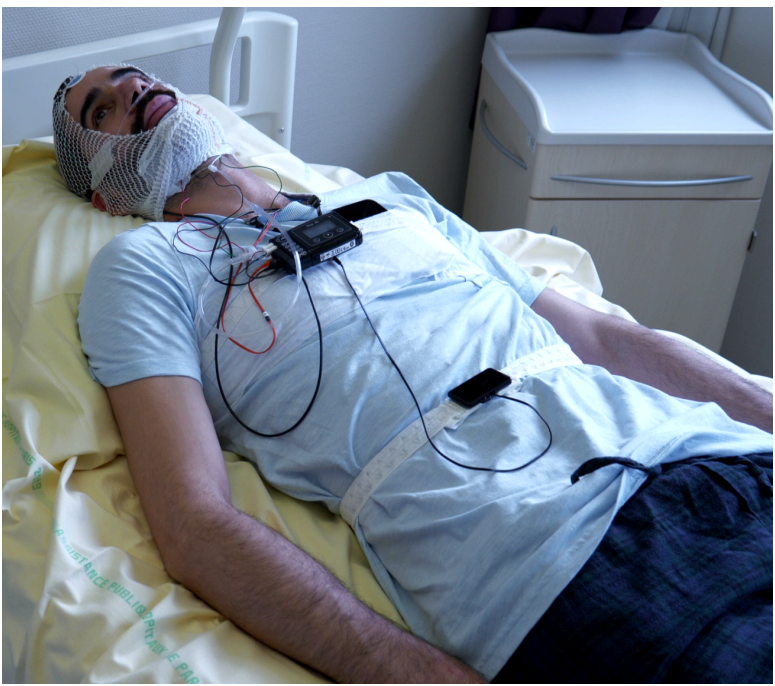

*Figure 1: full installation of a smartphone and the PSG on a patient*

The Apneal® application recorded the sound (8000 Hz) produced by the patients during the night, along with the 3D acceleration (100 Hz) and the 3D angular velocity (100 Hz) of the smartphone.

## Proprietary APNEAL algorithm

| Variable | Sample rate | Resolution |
|---|---|---|
| 3D acceleration, filtered using a bandpass filter between 0.1Hz and 1Hz | 100 Hz | $4.10^{-5}$g |
| Position (left, right, supine, prone or upright) | 0.033 Hz | - |
| Probability of breathing | 3.2 Hz | $3.10^{-5}$ |
| Snoring probability | 3.2 Hz | $3.10^{-5}$ |
| Audio | 8000 Hz | 16-bit |
| Audio power | 40 Hz | $4.10^{-3}$dB |
| Activity | 1Hz | $2.10^{-3}$ |
| ECG Heart rate | 1Hz | $2.10^{-3}$bpm |

*Table 1: Signals used by scorers*

Proprietary AI algorithms were applied to the raw signals recorded by the smartphones, to extract the position of the patients during the night, the probability of breathing, and the probability of the presence of snores. The sound (hearable by the scorers), the sound volume, the filtered acceleration on the three axes, and the activity were also extracted from the signals recorded by the smartphone.

The activity was extracted with the following method:
- Take the accelerometer L2-norm of the x and y axes
- Take the first derivate
- Take the absolute values of the first derivate
- Multiply by 10
- Clip to [0, 10] g/s

## Manual scoring of Apneal® recordings

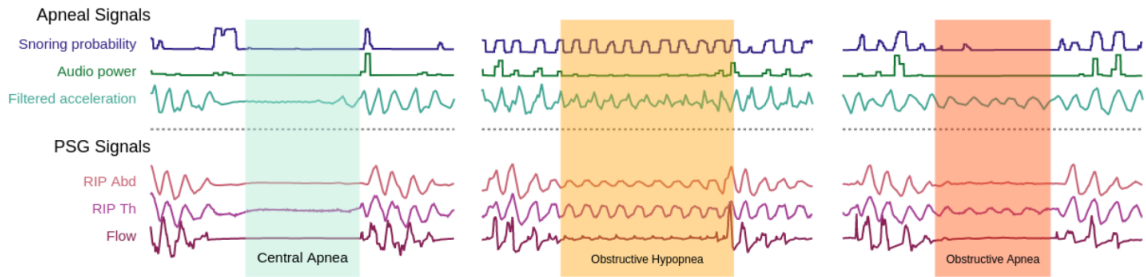

*Figure 2: Illustration of various respiratory events: central apnea, obstructive hypopnea, and obstructive apnea (from left to right), with, in parallel, some of the Apneal® signals used during manual scoring (top) and signals extracted from the PSG (bottom).*

The signals in Table 1 (all extracted from Apneal® recording apart from the ECG heart rate) were used by seven non-expert scorers to blindly score sleep stages (wake and sleep periods only) and respiratory events (central or obstructive apneas and hypopneas, and RERAs), following common guidelines. An illustration of the kind of signals they used can be seen during various respiratory events in Figure 2 (top) along with some of the signals recorded by the PSG during the same events (bottom). All scorers went through the following steps for each patient:
- Look for when the patient falls asleep for the first time and wakes up for the last time to set up the limits of the studied period
- Identify intra-sleep awakening periods (labeled as wake epochs if they are longer than 30 seconds)
- Identify arousals (less than 30 seconds long, and at least 3 seconds long)
- Once this is done for the whole studied period, go back to the beginning and look for respiratory events (wake epochs could also be added during this step)
- Score all the epochs not scored as wake periods as sleep periods.

Scorers were encouraged to listen to the audio when in doubt.

Before scoring, scorers received a one-hour training and scored one hour of recording to get feedback and correct their scoring process. Once trained, they all scored the same recording to study the consistency of scorings from one scorer to the other. After that, all the patients were scored by at least one scorer. The scorings produced by this process are considered as Apneal® manual scorings.

## Automatic scoring of respiratory events

In parallel, a sequential deep learning model was also used to automatically detect respiratory events during sleep periods, using input features extracted from Apneal® recordings. The sequential model was trained using cross-validation, with four folds of training, validation, and test sets, containing different patients' recordings. The scorings produced by this process are considered as Apneal® automatic scorings.

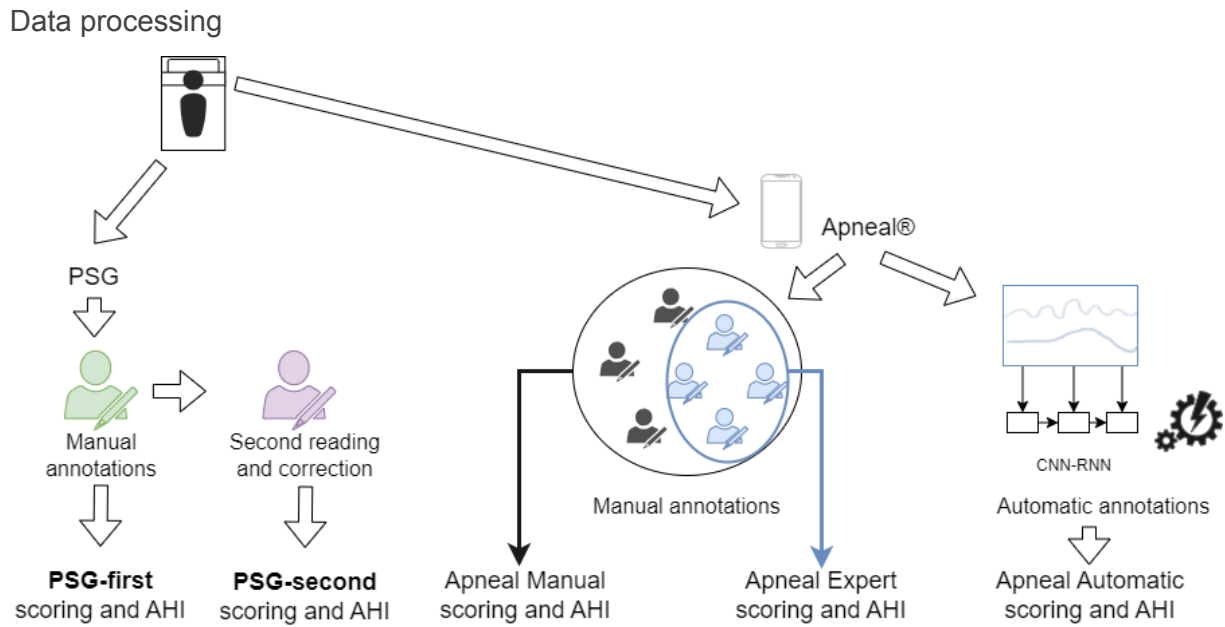

*Figure 3: Explanation of our data processing pipeline.*

After manual scoring, raw VPSG data on Alice 6 were extracted. After manual and automatic scorings of Apneal® recordings, the scorings of sleep stages and respiratory events of the different sources were extracted. See Figure 3 for a summary of our data processing pipeline.

For each patient and each setting, we extracted the Apnea and Hypopnea Index (AHI) found, which is the number of hypopneas and apneas during sleep periods divided by the number of hours of sleep. We compared the **PSG-first** AHI with the AHI obtained using the Apneal® manual, as they were done in a similar setting (one scorer per recording, one reading of the signal). We then compared **PSG-second** with Apneal® automatic scorings.

We also compared the respiratory events segmentation produced by these different methods.

Statistics

Patient characteristics were described as a median and interquartile range for quantitative variables and percentages for categorical variables.

We studied the ability of the Apneal® device to detect moderate-to-severe OSAS, i.e. classify patients under and above the AHI threshold of 15/h, the current threshold for SAS treatment (SPLF RPC 2009). For that, we computed positive predictive value and sensitivity. These values were obtained using the raw values of the predicted AHI and the threshold of value 15/h. To obtain 95% confidence intervals for these values, we used bootstrapping over patients, with N=10000.

We also computed the Area Under the ROC Curve (AUC-ROC) and the Area Under the Precision-Recall Curve (AUC-PR) for the identification of patients with an AHI above 15/h, and above 30/h. For the automatic scoring, we estimated the minimal threshold to set to obtain a precision of at least 0.9 and of at least 0.95 for the identification of patients above 15/h and above 30/h of AHI.

We computed the intraclass correlation coefficient (ICC) and Pearson correlation between PSG-scored AHI values and those obtained by the manual (with **PSG-first**) and automatic (with **PSG-second**) Apneal® scorings and provided the p-value for these values to assess the performances of the App.

We also compared the segmentation of respiratory events obtained from the PSG and the one obtained by the manual (with **PSG-first**) and automatic (with **PSG-second**) Apneal® scorings. We did that using positive predictive value and sensitivity metrics, considering a predicted event as a true positive when it overlaps with a PSG-based respiratory event for at least three seconds, with a margin for PSG-based respiratory events of 20 seconds. To obtain 95% confidence intervals for these values, we used bootstrapping over patients, with N=10000.

Ethical aspects

This study is part of the Evaluation of the Metrological Reliability of Connected Objects in the Measurement of Medical Physiological Parameters (EvalExplo) study [NCT03803098]. Ethics approval was obtained from Comité de Protection des Personnes Sud Est VI (approval number AU 1443), and written non-opposition was obtained according to the Jardé decree in France.

# RESULTS

Patient characteristics

A total of 46 patients were included. One patient was excluded because of a recording issue. Another patient was excluded because the phone placed on his chest was removed during the night, and put back in the wrong direction. The final sample included 44 patients with available PSG and Apneal® recordings. Patients' characteristics are described in Table 2.

| Age | Sex | BMI | N3 (min) | REM% | Sleep time (min) | AHI (PSG-first) | AHI (PSG-second) |
|---|---|---|---|---|---|---|---|
| med: 57.5, IQR: 18 | F 34% M 66% | med: 28.7, IQR: 7 | med: 78.8, IQR: 48.6 | med: 18.5% IQR: 8.7% | med: 371 IQR:82 | med 26.0 IQR: 31.1 | med 27.6 IQR: 32.1 |

Table 2: Patients' characteristics

Manual scoring

For classifying patients below and above AHI 15, the manual Apneal® scorings compared to the PSG scorings (**PSG-first**) lead to a sensitivity of 0.91 (95% CI [0.8, 1]), and a positive predictive value of 0.89 (CI 95% [0.76, 0.97]). For the AHI threshold of 30, we obtained a sensitivity of 0.85 (95% CI [0.67, 1]), and a positive predictive value of 0.94 (CI 95% [0.8, 1]).

Using varying thresholds on the predicted AHI we obtained an AUC-ROC of 0.85 (95% CI [0.69, 0.96]) and an AUC-PR of 0.94 (95% CI [0.84, 0.99]) for the identification of patients with an AHI above 15, and AUC-ROC of 0.95 (95% CI [0.86, 0.99]) and AUC-PR of 0.93 (95% CI [0.81, 0.99]) for the identification of patient with an AHI above 30.

The ICC between the AHI estimated manually, and the one obtained from the PSG scorings is 0.89 (p-value = $6.7 \times 10^{-17}$), Pearson correlation is 0.90 (p-value=$1.25 \times 10^{-17}$). Confusion matrix, regression plot, and Bland Altman plot can be seen in Figure 4. Results obtained at the event segmentation level are visible in Table 3.

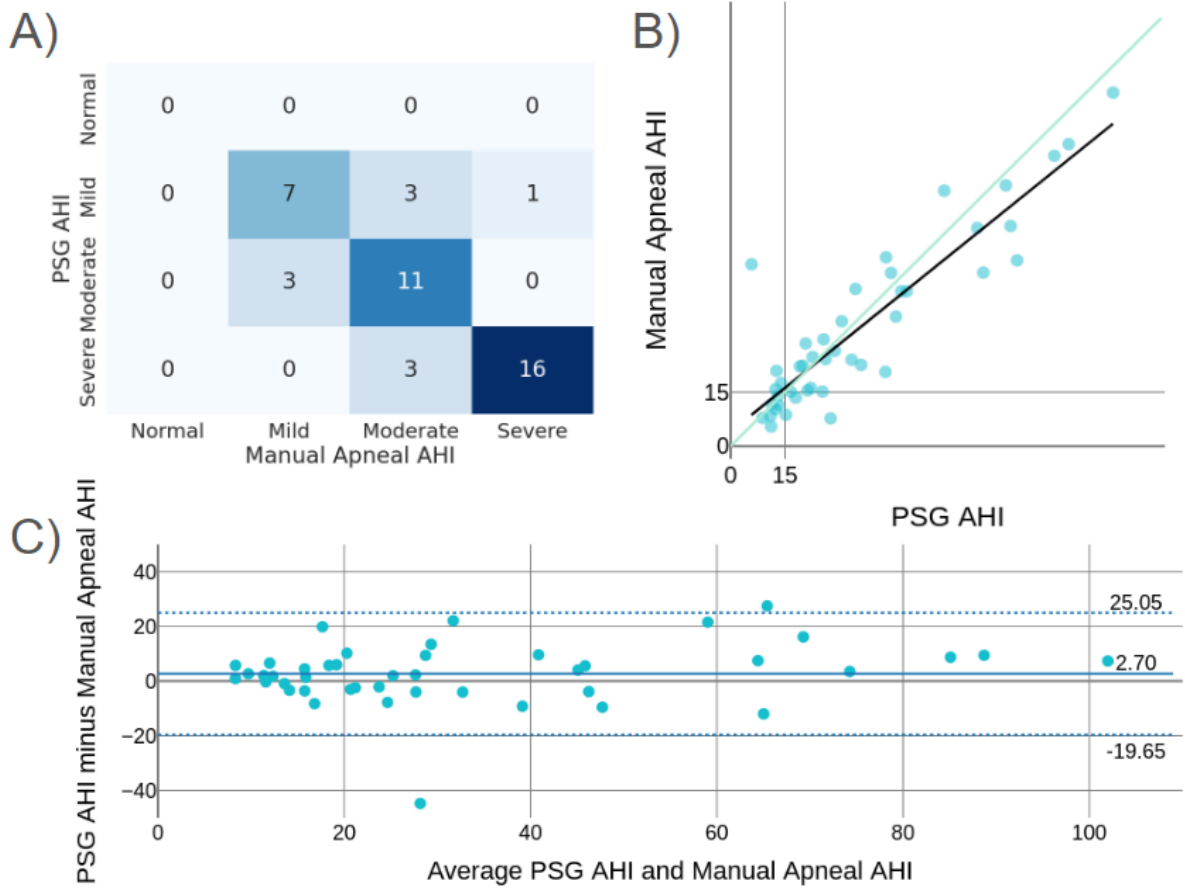

*Figure 4: Results of the manual scoring. A) Confusion matrix between the OSA severities obtained from the PSG AHI and the Manual Apneal® AHI. B) Regression plot between the PSG AHI and the automatic Apneal® AHI. C) Bland Altman plot between the PSG AHI and the Manual Apneal® AHI.*

Manual scoring by selected scorers

Using the recordings scored by all scorers, we were able to select the 4 best scorers on an independent test, who had the scorings that were the most consistent from one to the other and who were the closest to the PSG scoring for this recording.

We studied the results of Apneal® manual scorings for the subset of recordings that were scored by these selected scorers and we observed a sensitivity of 0.96 (95% CI [0.88, 1]) and a positive predictive value of 0.93 (95% CI [0.81, 1]) for the AHI threshold of 15. For the AHI threshold of 30, we obtained a sensitivity of 0.87 (95% CI [0.64, 1]), and a positive predictive value of 0.99 (CI 95% [0.95, 1]).

Using varying thresholds on the predicted AHI we obtained an AUC-ROC of 0.95 (95% CI [0.84, 1]) and an AUC-PR of 0.99 (95% CI [0.95, 1]) for the identification of patients with an AHI above 15, and AUC-ROC of 0.96 (95% CI [0.87, 1]) and AUC-PR of 0.96 (95% CI [0.86, 1]) for the identification of patients with an AHI above 30.

The ICC between the AHI estimated manually by selected scorers and the one obtained from the PSG (**PSG-first**) is 0.93 (p-value=2.6 x 10$^{-16}$), and the Pearson correlation is 0.95 (p-value = 1.8 x 10$^{-17}$).

Confusion matrix, regression plot, and Bland Altman plot can be seen in Figure 5. Results obtained at the event segmentation level are visible in Table 3.

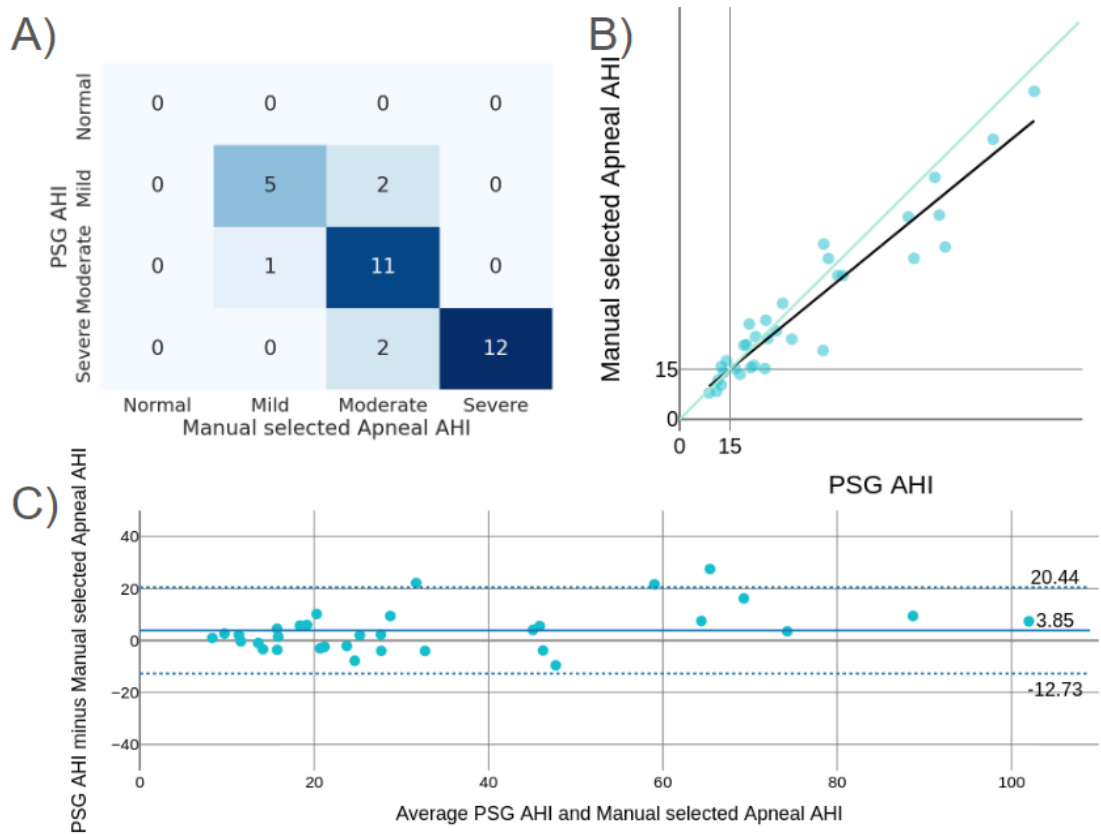

*Figure 5: Results of the manual scoring from selected scorers. A) Confusion matrix between the OSA severities obtained from the PSG AHI and the Manual selected Apneal® AHI. B) Regression plot between the PSG AHI and the Manual selected Apneal® AHI. C) Bland Altman plot between the PSG AHI and the Manual selected Apneal® AHI.*

Automatic scoring

For classifying patients below and above AHI 15, the automatic Apneal® scorings compared to the PSG scorings (**PSG-second**) lead to a sensitivity of 1 (95% CI [0.95, 1]), and a positive predictive value of 0.9 (95% CI [0.8, 0.9]) for the AHI threshold of 15. For the AHI threshold of 30, we obtained a sensitivity of 0.95 (95% CI [0.84, 1]), and a positive predictive value of 0.69 (CI 95% [0.52, 0.85]). The confusion matrix, regression plot, and Bland Altman plot can be seen in Figure 6.

Using varying thresholds on the predicted AHI we obtained an AUC-ROC of 0.85 (95% CI [0.64, 0.99]) and an AUC-PR of 0.97 (95% CI [0.9, 1]) for the identification of patients with an AHI above 15, and AUC-ROC of 0.87 (95% CI [0.74, 0.96]) and AUC-PR of 0.88 (95% CI [0.74, 0.96]) for the identification of patients with an AHI above 30.

We modified the thresholds used on the predicted AHI to identify patients' severity in order to obtain optimal predictive positive values. We obtained the updated following metrics. Using the smallest threshold to obtain a positive predictive value above 0.9: for 15 sensitivity of 0.97 (95% CI [0.91, 1]) and positive predictive value of 0.9 (95% CI [0.79, 0.98]); for 30 sensitivity of 0.57 (95% CI [0.35, 0.78]) and positive predictive value of 0.93 (95% CI [0.75, 1]).

The ICC between the AHI estimated, and the one obtained from the PSG scorings is 0.84 (p-value = 5.4 x $10^{-11}$) and the Pearson correlation found is 0.87 (p-value = 1.7 x $10^{-12}$).

Results obtained at the event segmentation level are visible in Table 3.

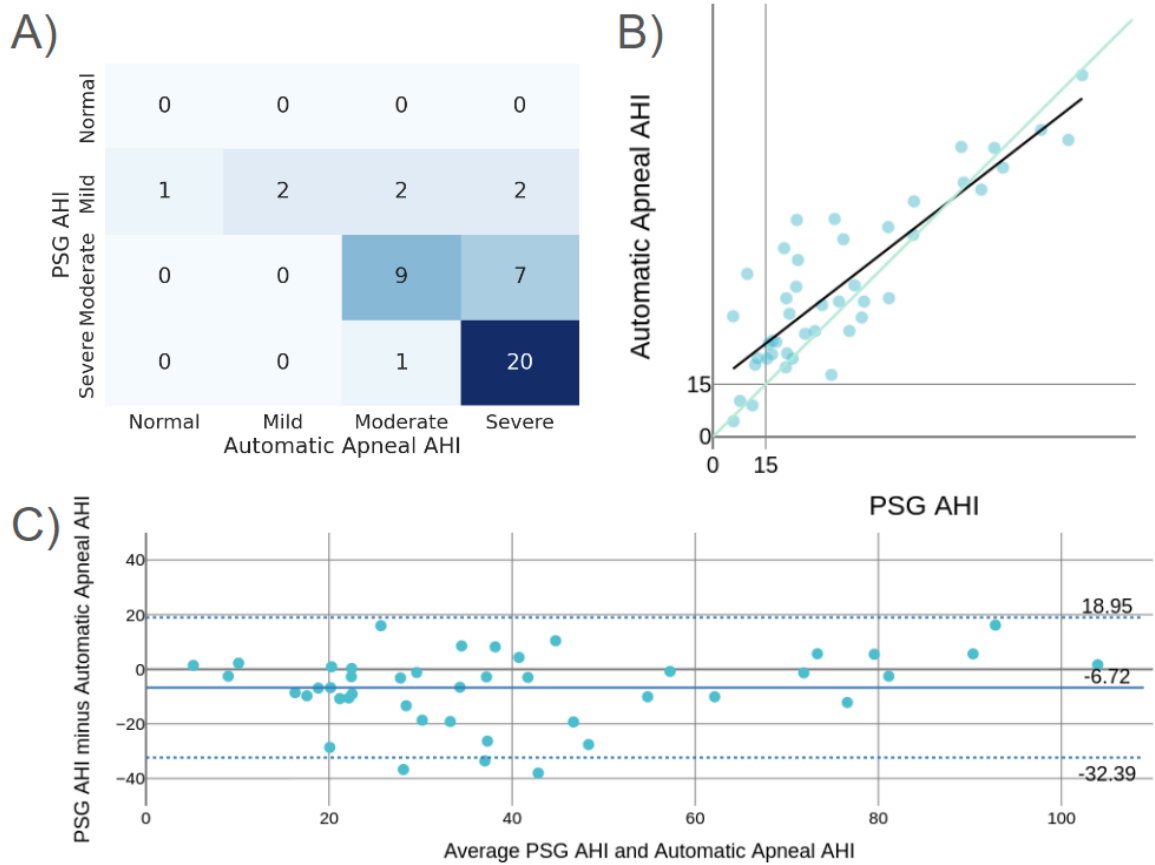

*Figure 6: Results of the automatic scoring. A) Confusion matrix between the OSA severities obtained from the **PSG-second** AHI and the automatic Apneal® AHI, using the severity threshold of IAH (< 5, 5-15 and > 15/h). B) Regression plot between the **PSG-second** AHI and the automatic Apneal® AHI. C) Bland Altman plot between the **PSG-second** AHI and the automatic Apneal® AHI.*

| Scoring type | PPV for segmentation | Sensitivity for segmentation |
|---|---|---|
| Manual | 0.68 [0.6, 0.74] | 0.7 [0.61, 0.77] |
| Manual Expert | 0.7 [0.62, 0.76] | 0.74 [0.66, 0.8] |
| Automatic | 0.69 [0.6, 0.76] | 0.77 [0.71, 0.82] |

Table 3: Event-per-event segmentation of respiratory events results for all scoring types produced from Apneal® recordings compared to PSG-based scorings (**PSG-first** for manual and **PSG-second** for automatic). The values here represent the ability of Apneal® to identify each individual respiratory event during a patient's night. The values displayed are median and 95% CI.

# DISCUSSION

This study presents the first steps for the validation of a new, minimally invasive tool to diagnose sleep apnea. Using only the smartphone's sensors (microphone, accelerometer, and gyroscope), respiratory events were detected properly. The signal recorded by smartphone sensors is enough for non-expert scorers to score respiratory events, and to detect apneic patients, with a sensitivity of 0.91, and a positive predictive value of 0.89 for the AHI threshold of 15. Moreover, this scoring could be automatized, using Apneal® version 0.1 automatic scoring of respiratory events obtaining a sensitivity of 1 and a PPV of 0.9. The predicted AHI values seem to be relevant to identifying the SAHS severity of patients (AHI threshold of 15 and 30, the only ones that can be tested considering the AHI distribution of patients that were included in this study), as we obtained an AUC-ROC of 0.95 (threshold 15) and 0.96 (threshold 30) for the manual scoring (selected scorers) and an AUC-ROC of 0.85 (threshold 15) and 0.87 (threshold 30) for the automatic scoring. These results are encouraging and provide us with a proof of concept of this method.

Furthermore, interestingly, this method does not just provide with a global AHI but enables us to segment individual respiratory events accurately (see Table 3), as we show that manual scoring can reach up to 0.7 of PPV and 0.74 of sensitivity for individual respiratory events segmentation, and the automatic scoring 0.69 of PPV and 0.77 of sensitivity. Further work on the algorithm is necessary to increase the global performance of the solution.

Compared to existing solutions, our AI-driven solution shows good performance in screening and diagnosis. Indeed, screening relies on questionnaires such as the STOP-BANG, Berlin Questionnaire. Their sensitivity and specificity are lower than 80% in the general population (Abrishami et al., 2010; Chung & Vairavanathan, 2008), and they are not adapted to specific populations (e.g pregnant women, children, psychiatry…). Epworth Sleepiness Scale (ESS) is intended to detect excessive daytime sleepiness and is often used although not recommended as a screening tool. Our device, although it would need specific validation in children and pregnant women, represents an easily accessible tool for SAHS objective detection. Other currently available AI-driven solutions include pulse tonometry and mandibular movements. Their diagnostic performances are equivalent to the ones we find in this preliminary study[11,18].

As far as the cost of diagnosis is concerned, this has been considerably reduced by the systematic introduction of home ventilatory polygraphy, which can be performed in private practice by doctors of various specialties (pulmonologist, ENT specialist, cardiologist, psychiatrist) or by the general practitioner specializing in sleep[19]. Polygraphy involves a limited number of sensors, enabling patients to equip themselves independently and carry out the recording at home. Although one night's hospitalization is avoided and the cost is, therefore, lower [20], the investment in equipment and the logistics of the examination (loan of equipment, recovery, disinfection, reading of tracings) remain a barrier to scaling up healthcare systems diagnostics capacity. New devices using derivative signals and AI, such as jaw movements (Sunrise®), ballistocardiography embedded in a device placed under the mattress (Withings®) or arterial pulse tonometry (Watchpat®) have shown good performance for sleep apnea detection. In addition, it has been demonstrated with such devices that AHI may vary considerably from one night to another, leading to consider the need to record for several nights to set a proper diagnosis[13]. Such a consideration further pinpoints the need for simplified devices for SAHS diagnosis. However, the later methods still require a dedicated device, inducing the need for an in-person visit (at least to a pharmacy or a healthcare professional) to set up the exam and/or retrieve results, but also addresses the question of device recycling or elimination. Our device, using only the patient's smartphone, ensures access to sleep apnea diagnosis in remote areas and has a lower carbon footprint.

Strengths and limits of the study

Strengths of our study include the blinded manual scoring to assess the quality of the recorded signals, the use of polysomnography and not polygraphy as a gold standard, and the high number of respiratory events to be detected, thanks to the inclusion of patients with a high probability of sleep apnea.

One of the limitations of our work is the use of the heart rate provided by the ECG of the PSG during manual scorings. The algorithm to extract the heart rate from the accelerometer and the gyroscope was developed subsequently, and thus could not be used for this first step. However, articles on the extraction of heart rate from the seismocardiogram show that an equivalent heart rate can be measured using that type of signal [21]. This would be enough to reproduce these results using only signals extracted from the smartphone. This error reduces to 3.8 bpm during movement-free and artifact-free periods.

Outliers in the results figures for the automatic scoring (Figure 6) are visible: the automatic scoring tends to overestimate the number of respiratory events for some patients. These outliers may be due to multiple factors, including the confusion of periodic leg movements with movements due to respiratory recovery, and thus respiratory events like apneas or hypopneas. The distinction between these two types of events will be taken care of by future versions of Apneal® automatic scoring to solve this overestimation issue.

Another limitation is that we are not using the same gold standard scoring to evaluate our manual and automatic scorings. To evaluate Apneal® manual scoring, we use **PSG-first**, a scoring that was made (manually) based on the PSG by a first specialized doctor or technologist. To evaluate Apneal® automatic scoring, we use **PSG-second**, a scoring that is the clean and verified version of **PSG-first**, which was checked by another sleep technologist. We want to underline that these two scorings could be used as a reference. However, we chose this evaluation setting to have a fairer comparison. Indeed, the manual scoring was made in the same settings as **PSG-first**, with the same risk of missing respiratory events during the scoring. As the application of a model does not lead to attention issues (a model misses an event only if it makes a mistake, not because its attention span), we wanted to compare the automatic scoring to a more "perfect" reference scoring.

Last, our sample is relatively small. The monocentric set-up in a reference sleep center allows for a proper gold standard with highly trained technicians and sleep doctors for scoring but induces a selection bias by including patients with severe sleep complaints. Including subjects with a low pretest probability is mandatory in the next steps to ensure proper device validation in a less specific population. As this paper showed the feasibility of using only smartphone sensors to help diagnose sleep apnea, the next steps are to improve the automatic scoring of respiratory events using these signals and to generalize this method on more and diverse patients. We plan to achieve a multicentric clinical study of 500 patients to validate the method.

CONCLUSION

This work introduces a proof of concept demonstrating the potential of smartphone-based recordings of various signals for detecting respiratory events associated with SAHS diagnosis. Our findings indicate that manual scoring of these signals is possible and accurate compared to PSG-based scorings, demonstrating the interpretability of the recorded signals. Additionally, we also present the first version of an automatic scoring method based on a deep learning model, which provides promising results. A larger multicentric validation study, involving subjects with different SAHS severity is required to confirm these results. Further work will also be done to improve the performances of SAHS severity identification in future versions of the automatic scoring algorithm.

# ACKNOWLEDGEMENTS

The authors are thankful to the patients who agreed to participate.

The authors are thankful to the sleep technicians (Rémi Cellot, Béatrice Guy, Marie-Cécile Flottes, Carine Ecourtemer, Axelle Lefebvre-Roque, Rosine Zana, Claire Petrovic) and to Fedja Kerzabi for managing the Apneal® application and smartphones.

The authors are thankful for manual scoring, to Danica Despotović, Rima El Kosseifi, Séverin Benizri, Anton Prodanet, employees of Mitral SAS.

Funding: Mitral MSD France (unrestricted grant to the Assistance Publique Hôpitaux de Paris Foundation). This research is partially supported by the Agence Nationale de la Recherche as part of the "Investissements d'avenir" program (reference ANR-19-P3IA-0001; PRAIRIE 3IA Institute).

This research is partially supported by the Agence Nationale de la Recherche as part of the "Investissements d'avenir" program (reference ANR-19- P3IA-0001; PRAIRIE 3IA Institute).

Conflicts of interest:

Guillaume Cathelain and Juliette Millet are both employees of Mitral.